\documentclass[3p,number]{elsarticle}
\makeatletter
\def\ps@pprintTitle{%
	 \let\@oddhead\@empty
	 \let\@evenhead\@empty
     \def\@oddfoot{\footnotesize\itshape
	   Preprint \ifx\@journal\@empty
	   \else\@journal\fi\hfill\today}%
	 \let\@evenfoot\@oddfoot}
\makeatother
\usepackage{hyperref}
\usepackage{etex}
\usepackage{amsmath}
\usepackage{amssymb}
\usepackage{amsthm}
\usepackage{array}
\usepackage{booktabs}
\usepackage{graphicx}
\usepackage{calc}
\usepackage{url}
\usepackage{multirow}
\usepackage{wrapfig}
\usepackage{tikz}
\usetikzlibrary{arrows,automata,shapes,shapes.multipart,decorations.markings,positioning}
\usepackage[all]{xy}
\usepackage[trim]{tokenizer}
\usepackage[super]{nth}
\usepackage{complexity}
\usepackage[labelformat=simple]{subcaption}

\newcommand{\outputs}{\mathsf{outputs}}
\newcommand{\notsorted}{\mathsf{notsorted}}

\newcommand{\sset}[2]{\left\{~#1  \left|
      \begin{array}{l}#2\end{array}
    \right.     \right\}}

\newcounter{sncolumncounter}
\newcounter{snrowcounter}

\newcommand{\Set}[2]{%
	\{\, #1 \mid #2 \, \}%
}

\newcommand{\sncolwidth}{0.7} %

\def \nodeconnection#1{%
	\foreach \i in {#1}{%
		\GetTokens{nodesrc}{nodedest}{\i}
		\draw (\value{sncolumncounter}*0.7,\value{snrowcounter}-\nodesrc) node[circle,inner sep=0pt,minimum size=3pt,fill=black]{}--(\value{sncolumncounter}*0.7,\value{snrowcounter}-\nodedest) node[circle,inner sep=0pt,minimum size=3pt,fill=black]{};
	}
	\addtocounter{sncolumncounter}{1}
}
\newenvironment{sortingnetwork}[2]
{
  \setcounter{sncolumncounter}{0}
  \setcounter{snrowcounter}{#1}
  \def \sn@fullsize{15}
  \begin{tikzpicture}[scale=#2*0.7]
}
{
  \foreach \i in {1, ..., \value{snrowcounter}}
  {
    \draw (-\sncolwidth,\i)--(\sncolwidth*\value{sncolumncounter}+\sncolwidth,\i);
  }
  \end{tikzpicture}
}
\makeatother

\begin{document}

\title{SAT encodings for sorting networks, single-exception sorting networks and $\epsilon-$halvers.\tnoteref{t1}} 

\tnotetext[t1]{Supported by the Spanish MINECO project TEC2015-69266-P (FEDER, UE)}  

\author{Jos\'e A. R. Fonollosa} \ead{jose.fonollosa@upc.edu} 

\address{Department of Signal Theory and Communications, Universitat Polit\`ecnica de Catalunya, Barcelona, Spain}

\begin{abstract}
Sorting networks are oblivious sorting algorithms with many practical applications and rich theoretical properties. Propositional encodings of sorting networks are a key tool for proving concrete bounds on the minimum number of comparators or depth (number of parallel steps) of sorting networks. In this paper, we present new SAT encodings that reduce the number of variables and clauses of the sorting constraint of optimality problems. Moreover, the proposed SAT encodings can be applied to a broader class of problems, such as the search of optimal single-exception sorting networks and $\epsilon-$halvers. We obtain optimality results for single-exception sorting networks on $n \le 10$ inputs.
\end{abstract}

\maketitle

\section{Introduction}
\label{sec:intro}
A sorting algorithm is data-independent or oblivious if the sequence of comparisons does not depend on the input list. Sorting networks are oblivious sorting algorithms with many practical applications and rich theoretical properties \cite{knuth1998art}. From the practical point of view, sorting networks are the usual choice for simple parallel implementations in both hardware and software such as Graphics Processing Units (GPUs). Moreover, sorting networks are also of interest for secure computing methods like secure multi-party computation, circuit garbling and homomorphic encryption \cite{bogdanov2014practical}. Other applications include median filtering, switching circuits, and encoding cardinality constraints in propositional satisfiability problems (SAT)\cite{abio2013parametric}. Interestingly, we use this cardinality constraint in \cite{fonollosa2018joint} to obtain optimal sorting networks and here to search optimal single-exception sorting networks.

From the theoretical point of view comparator networks can be studied using the combinatorial and algebraic properties of permutations \cite{bruijn1974,bruijn1983}, as well as constrained boolean monotone circuits using the zero-one principle \cite[p.~223]{knuth1998art}. In the usual representation, the $n$ input values are fed into networks of $n$ channels connected by comparators that swap unordered inputs from two channels. The sequence of data-independent comparisons can be parallelized grouping independent comparators in layers. The depth of a comparator network is the number of layers, i.e., the delay in a parallel implementation.

\begin{figure}[htb]
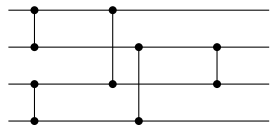

	\centering
	\begin{sortingnetwork}{4}{0.7}
		\nodeconnection{ {0, 1}, {2, 3}}
		\addtocounter{sncolumncounter}{2}
		\nodeconnection{ {0, 2}}
		\nodeconnection{ {1, 3}}
		\addtocounter{sncolumncounter}{2}
		\nodeconnection{ {1, 2}}
	\end{sortingnetwork}
	\caption{Comparator network of depth $3$ with $5$ comparators.}
	\label{fig:example}
\end{figure}

The typical graphical representation of a comparator network is depicted in Figure \ref{fig:example}.

SAT encodings of sorting networks has been recently used to obtain new optimal-size \cite{codish2016nine} and optimal-depth sorting networks \cite{bundala2014optimal-depth, ehlers2015new, codish2016sorting}, as well as joint size and depth optimality results \cite{fonollosa2018joint}

In this paper we propose new SAT encodings of sorting networks that assigns a variable to each possible input or output vector after each comparator. In this encoding framework, a sorting networks is a network in which the set of unsorted outputs is empty, or that sorts of all its inputs. Moreover, the propositional encoding based on the set of unsorted inputs can also be used to characterize networks that sorts all zero-one inputs (bit-strings) except one (called \textit{single-exception sorting networks}) or merging networks. While the propositional encoding based on the set of unsorted outputs can be used to characterize perfect halvers and $\epsilon-$halvers \cite{ajtai1983sorting}.

Single-exception sorting networks has been studied by Chung and Ravikumar \cite{chung1987size, chung1989strong}, and Parberry \cite{parberry1990single,parberry1991computational} as the key component of the proof that the sorting network verification problem is $\coNP$ complete. In \cite{parberry1991computational} Parberry conjectured that $D_1(n)$, the minimum-depth of an $n-$channel single-exception sorting network was equal to $D(n)$, the minimum depth of a $n-$channel sorting network. We show that the conjecture is true for $n = 4$ and $6 \le n \le 10$. However, $D_1(n) = D(n) - 1$ for $n = 5$ and the trivial cases $n = 2$ and $n = 3$. We also study the minimum size of single-exception sorting networks.

A (perfect) halver on $n=2m$ channels is a comparator network that split the input vector in 2 blocks. At the output, the $m$ smallest inputs are in the first $m$ channels and the $m$ largest inputs in the other channels. Perfect halvers must have a depth greater than $\log_2(m)$. However, there are approximate halvers ($\epsilon-$halvers) of constant-depth (dependent on the approximation factor $\epsilon$ but not in the number of channels $n$). $\epsilon-$halvers are important comparator networks because they are the basic blocks of the asymptotically optimal AKS sorting network \cite{ajtai1983sorting} and more recent variants such as \cite{goodrich2014zig}.

\section{Preliminaries}
\label{sec:Preliminaries}
A comparator network $C$ is a set of channels connected by a sequence of comparators as illustrated in Figure \ref{fig:example}. Channels are depicted as horizontal lines (with the first channel at the top). Each comparator (i,j) compares the input values ($in_i$, $in_j$) of the two connected channels $(1 \le i < j \le n)$ and if necessary rearrange them such that $out_i=min(in_i, in_j)$ and $out_j=max(in_i, in_j)$. The sequence of comparators can be grouped in maximal sets of independent comparators (layers) whose output can be computed in parallel. The depth of a comparator network is the number of layers. A sorting network is a comparator network that sorts all input sequences.

A key tool for the proof of correctness of sorting networks is the 0-1-principle \cite{knuth1998art}: if a sorting network for n channels sorts all $2^n$ sequences of 0's and 1's, then it sorts every arbitrary sequence of values.

Let $C$ be a comparator network, $x=(x_1\ldots x_n)\in\{0,1\}^n$ an input vector, and $v^k=(v^k_1\ldots v^k_n)$ the output of the network after layer $k$. The value $v^k_i$ carried by channel $i$ after layer $k$ propagates through $C$ as follows. $v^0_i = x_i$, and for $0 < k \le d$:

\[v^k_i=\begin{cases}
min(v^{k-1}_i,v^{k-1}_j) & \mbox{if there is a comparator between channels $i$ and $j > i$} \\
max(v^{k-1}_j,v^{k-1}_i) & \mbox{if there is a comparator between channels $j$ and $i > j$} \\
v^{k-1}_i & \mbox{otherwise}
\end{cases}\]

The output of the network for input $x$ is $C(x)=v^d$, and $\outputs(C)=\sset{C(x)}{x\in\{0,1\}^n}$. The comparator network $C$ is a \emph{sorting network} if all elements of $\outputs(C)$ are sorted. A comparator network does not change the input values (the number of 1's and 0's). Hence, the minimum cardinality of $\outputs(C)$ in $n + 1$, and a comparator networks is a sorting network if and only if it achieves this cardinality.

An $\epsilon-$halver is a comparator network on $n=2m$ channels such that, for any $k \le m$, at most $\epsilon k$ of the largest $k$ inputs will be in the upper half of the output and at most $\epsilon k$ of the smallest $k$ inputs will be in the lower half of the output, where $\epsilon \ge 0$.

\section{Propositional encodings for fixed-size comparator networks}
\label{sec:SATs}

In this section we derive two new SAT encodings of interest for the optimal-size problem of sorting networks and other comparator networks.

A comparator network $C=(c_1\ldots c_s)$ of size $s$ on $n$ channels is a sequence of $s$ comparators represented by a set of Boolean variables $C^s_n=\sset{g^k_{i,j}}{1 \leq i < j \leq n,1 \leq k \leq s}$, the value of $g^k_{i,j}$ indicating if the $k$\textsuperscript{th} comparator connects channels $i$ and $j$, i.e., $c_k=(i,j)$. A comparator formed by a contiguous subsequence of the comparators in $C$ is denoted as $C^{a:b}=(c_a\ldots c_b)$

\subsection{Validity encodings}
A valid network with $s$ comparators has only one comparator for each $k$. We can use any one-hot encoding over $g^k_{i,j}$ for each $k$.

\begin{align*}
\mathit{AtMostOneSize}^k(C^s_n) =& \bigwedge_{1 \leq i < j \leq n, i < l < m \leq n} (\neg g^k_{i,j} \vee \neg g^k_{l,m}) \\
\mathit{AtLeastOneSize}^k(C^s_n) =& \bigvee_{1 \leq i < j \leq n} g^k_{i,j} \\
\mathit{ValidSize}(C^s_n) =& \bigwedge_{1 \leq k \leq n} \mathit{AtMostOneSize}^k(C^s_n) \wedge \mathit{AtLeastOneSize}^k(C^s_n)
\end{align*}

In the following subsections we present two alternative sorting constraints. The first one is based on encoding $\outputs^k(C)=\sset{C^{1:k}(x)}{x\in\{0,1\}^n}$, the set of output vectors after the $k$\textsuperscript{th} comparator.

\subsection{Fixed-size forward encoding (SFWD)}
\label{subsec:forward}
We encode each possible vector $v^k=(v^k_1\ldots v^k_n)$ at the output of comparator $k$ with a Boolean variable $o^k_m$, with $0 \le k \le s, 0 \le m < 2^n$, where $m$ is the integer with binary representation $v^k$, with $v^k_1$ the least significant bit and $v^k_n$ the most significant bit. Let the expression $m = c_{i,j}(w)$, with $0 \le m, w < 2^n$, denote that a comparator ($i$, $j$) transforms a vector with the binary representation of the integer $w$ into the binary representation of the integer $m$. And $m = sorted(w)$ that the binary representation of $m$ is the sorted version of the binary representation of $w$.

The set $\outputs^k(C)$ is defined by the variables $o^k_m$ indicating if the corresponding binary representation of $m$ is an element of that set. We can now encode the relation between the vectors in $\outputs^k(C)$ and the vectors at the output of the previous comparator, $\outputs^{k-1}(C)$, as follows:

\begin{align*}
\mathit{FwdUpdate}^k_m(C^s_n) = &\bigwedge_{1 \leq i < j \leq n} \left( g^k_{i,j} \rightarrow \mathit{Fwd}^k_{i,j,m} \right) \\
\mathit{ForwardSize}(C^s_n) = &\bigwedge_{1 \leq k \leq s, 0 \le m < 2^n} \mathit{FwdUpdate}^k_m(C^s_n)
\end{align*}

where

\[
\mathit{Fwd}^k_{i,j,m}=\begin{cases}
o^k_m \leftrightarrow o^{k-1}_m \vee o^{k-1}_w & \mbox{if $\exists w \ne m$, such that $m = c_{i,j}(w)$} \\
o^k_m \leftrightarrow o^{k-1}_m                & \mbox{if $m = c_{i,j}(m)$ and $\nexists w \ne m$, such that $m = c_{i,j}(w)$} \\
\neg o^k_m                                     & \mbox{if $m \ne c_{i,j}(m)$ and $\nexists w \ne m$, such that $m = c_{i,j}(w)$}
\end{cases}
\]

The $\mathit{FwdUpdate}^k_m$ constraint describes the impact of each comparator on the Boolean variable $o^k_m$, and the $\mathit{ForwardSize}$ equation includes the $\mathit{FwdUpdate}$ constraints for all the vectors and comparators in the network.

A sorting network for $n$ channels with $s$ comparators exists if and only if there is a solution in which $\outputs(C)=\outputs^s(C)$ does not contain any unsorted vector:

\begin{equation}\label{eq:sat_f}
\varphi^f_s(n,s) = \mathit{ValidSize}(C^s_n) \wedge \mathit{ForwardSize}(C^s_n) \wedge \mathit{AllInputs}_n \wedge \mathit{NoUnsortedOutputs}^s_n
\end{equation}

with

\begin{align*}
\mathit{AllInputs}_n =              & \bigwedge_{0 \le m < 2^n} o^0_m \\
\mathit{NoUnsortedOutputs}^s_n = & \bigwedge_{m \ne sorted(m)} \neg o^s_m
\end{align*}

\subsection{Fixed-size backward encoding (SBCK)}

We can also analyze the behavior of a comparator network studying the set of inputs that are not sorted by the network, i.e., the set $\notsorted(C)=\Set{x}{C(x) \ne sorted(x)}$. The Boolean variable $q^k_m$ is used in this case to indicate if the corresponding vector is an element of $\notsorted^k(C)=\Set{x}{C^{k+1:s}(x) \ne sorted(x)}$, the set of inputs that are not sorted by the last $s-k$ comparators of the network. We can now relate the vectors in $\notsorted^{k-1}(C)$ with the vectors in $\notsorted^{k}(C)$ as follows:

\begin{align*}
\mathit{Bck}^k_{i,j,m}            &= q^{k-1}_m \leftrightarrow q^k_w \mbox{, with $w = c_{i,j}(m)$} \\
\mathit{BckUpdate}^k_m(C^s_n)     &= \bigwedge_{1 \leq i < j \leq n} \left( g^k_{i,j} \rightarrow \mathit{Bck}^k_{i,j,m} \right) \\
\mathit{BackwardSize}(C^s_n)      &= \bigwedge_{1 \leq k \leq s, 0 \le m < 2^n} \mathit{BckUpdate}^k_m(C^s_n)
\end{align*}

In this case, a sorting network for $n$ channels with $s$ comparators exists if and only if there is a solution in which all the (unsorted) inputs are sorted, i.e., if $\notsorted(C)=\notsorted^{0}(C)$ is empty:

\begin{equation}\label{eq:sat_b}
\varphi^b_s(n,s) = \mathit{ValidSize}(C^s_n) \wedge \mathit{BackwardSize}(C^s_n) \wedge \mathit{Outputs}^s_n \wedge \mathit{NoUnsortedInputs}_n
\end{equation}

with

\begin{align*}
\mathit{Outputs}^s_n =        & \bigwedge_{m \ne sorted(m)} q^s_m \wedge \bigwedge_{m = sorted(m)} \neg q^s_m \\
\mathit{NoUnsortedInputs}_n = & \bigwedge_{0 \le m < 2^n} \neg q^0_m
\end{align*}

The backward encoding is also useful for single-exception sorting networks, since we can easily encode the single-exception constraint with any one-hot encoding of $q^0_m$:

\begin{align*}
\mathit{AtMostOneUnsorted}_n =   & \bigwedge_{0 \leq m < w < 2^n} (\neg q^0_m \vee \neg q^0_w) \\
\mathit{AtLeastOneUnsorted}_n =  & \bigvee_{0 \leq m < 2^n} q^0_m \\
\mathit{SingleUnsortedInput}_n = & \mathit{AtMostOneUnsorted}_n \wedge \mathit{AtLeastOneUnsorted}_n
\end{align*}

The resulting encoding of fixed-size single-exception networks is:

\begin{equation}\label{eq:sat1}
\varphi^1_s(n,s) = \mathit{ValidSize}(C^s_n) \wedge \mathit{BackwardSize}(C^s_n) \wedge \mathit{Outputs}^s_n \wedge \mathit{SingleUnsortedInput}_n
\end{equation}

\section{Propositional encodings for fixed-depth comparator networks}
\label{sec:SATd}

In this section we adapt the previous results to derive SAT encodings for fixed-depth comparator networks. We fix the number of layers to $d$, and the comparators are represented by a set of Boolean variables $C^d_n=\sset{g^k_{i,j}}{1 \leq i < j \leq n,1 \leq k \leq d}$.

\subsection{Validity encodings}
In a valid network the comparators of each layer are independent, i.e., each channel may be used only once:

\begin{align*}
\mathit{AtMostOneDepth}^k_i(C^d_n) = & \bigwedge_{1 \leq i \neq j \neq l \leq n} \left( \neg g^k_{\min(i, j), \max(i, j) } \vee \neg g^k_{\min(i, l), \max(i, l) } \right) \\
\mathit{ValidDepth}(C^d_n) =         & \bigwedge_{1 \leq k \leq d, 1 \leq i \leq n} \mathit{AtMostOneDepth}^k_i(C^d_n)
\end{align*}

\subsection{Fixed-depth forward encoding (DFWD)}

We divide each layer in $n-1$ sublayers with at most one comparator in each of them. Then, we apply the forward coding of subsection \ref{subsec:forward} to each sublayer. Each possible binary vector at the output of the sublayer $i$ of layer $k$ ($\outputs^{k,i}(C)$) is represented with a Boolean variable $p^{k,i}_m$, with $0 \le k \le d, 1 \le i < n, 0 \le m < 2^n$, and we propagate $p^{k,i}_m$ sublayer by sublayer. For each layer $k$, the sublayer $i$ contains the comparator connecting channel $i$ with another channel $j > i$, or is empty if there is not such comparator in that layer.

\begin{align*}
\mathit{Sublayer}^{k,i}(C^d_n) =            & \bigvee_{i < j \leq n} g^k_{i,j} \\
\mathit{FwdSublayerUpdate}^{k,i}_m(C^d_n) = & \left( \neg \mathit{Sublayer}^{k,i}(C^d_n) \rightarrow \mathit{Fwd0}^{k,i}_{m} \right) \wedge
                                              \bigwedge_{i < j \leq n} \left( g^k_{i,j} \rightarrow \mathit{Fwd}^{k,i}_{j,m} \right) \\
\mathit{ForwardDepth}(C^d_n) =              & \bigwedge_{1 \leq k \leq d, 1 \leq i < n, 0 \le m < 2^n} \mathit{FwdSublayerUpdate}^{k,i}_m(C^d_n) 
\end{align*}

where

\[
\mathit{Fwd0}^{k,i}_{m}=\begin{cases}
p^{k,i}_m \leftrightarrow p^{k-1,n-1}_m & \mbox{if $i=1$} \\
p^{k,i}_m \leftrightarrow p^{k,i-1}_m   & \mbox{otherwise}
\end{cases}
\]

and

\[
\mathit{Fwd}^{k,i}_{j,m}=\begin{cases}
p^{k,i}_m \leftrightarrow p^{k-1,n-1}_m \vee p^{k-1,n-1}_w & \mbox{if $i=1$ and $\exists w \ne m$, such that $m = c_{i,j}(w)$} \\
p^{k,i}_m \leftrightarrow p^{k-1,n-1}_m                    & \mbox{if $i=1$ and $m = c_{i,j}(m)$ and $\nexists w \ne m$, such that $m = c_{i,j}(w)$} \\
p^{k,i}_m \leftrightarrow p^{k,i-1}_m \vee p^{k,i-1}_w     & \mbox{if $i>1$ and $\exists w \ne m$, such that $m = c_{i,j}(w)$} \\
p^{k,i}_m \leftrightarrow p^{k,i-1}_m                      & \mbox{if $i>1$ and $m = c_{i,j}(m)$ and $\nexists w \ne m$, such that $m = c_{i,j}(w)$} \\
\neg p^{k,i}_m                                             & \mbox{if $m \ne c_{i,j}(m)$ and $\nexists_{w \ne m}$, such that $m = c_{i,j}(w)$}
\end{cases}
\]

A sorting network for $n$ channels with $d$ layers exists if and only if there is a solution in which $\outputs(C)=\outputs^{d,n-1}(C)$ does not contain any unsorted vector:
\begin{equation}\label{eq:sorts_forward_depth}
\varphi^f_d(n,d) = \mathit{ValidDepth}(C^d_n) \wedge \mathit{ForwardDepth}(C^d_n) \wedge \mathit{AllInputs}_n \wedge \mathit{NoUnsortedOutputs}^d_n
\end{equation}

with

\begin{align*}
\mathit{AllInputs}_n =              & \bigwedge_{0 \le m < 2^n} p^{0,n-1}_m \\
\mathit{NoUnsortedOutputs}^d_n = & \bigwedge_{m \ne sorted(m)} \neg p^{d,n-1}_m
\end{align*}

In this encoding framework, we can easily consider other comparator networks defined in terms of valid outputs such as halvers and $\epsilon-$halvers. We just need to replace the ${m \ne sorted(m)}$ index selection in the $\mathit{NoUnsortedOutputs}^d_n$ equation with a generic ${invalid(m)}$ that forbids invalid outputs.

\subsection{Fixed-depth backward encoding (DBCK)}

We can use the same sublayers idea to derive the fixed-depth version of the backward encoding from the fixed-size backward equations. The Boolean variable $r^{k,i}_m$, with $0 \le k \le d, 1 \le i < n, 0 \le m < 2^n$ indicates if the binary representation of $m$ is an element of $\notsorted^{k,i}(C)$, the set of vectors that are not sorted by the sequence of comparators after the sublayer $i$ of layer $k$. The equations that relate each sublayer are: 

\begin{align*}
\mathit{BckSublayerUpdate}^{k,i}_m(C^d_n) = & \left( \neg \mathit{Sublayer}^{k,i}(C^d_n) \rightarrow \mathit{Bck0}^{k,i}_{m} \right) \wedge
                                           \bigwedge_{i < j \leq n} \left( g^k_{i,j} \rightarrow \mathit{Bck}^{k,i}_{j,m} \right) \\
\mathit{BackwardDepth}(C^d_n) = &\bigwedge_{1 \leq k \leq s, 1 \leq i < n, 0 \le m < 2^n} \mathit{BckSublayerUpdate}^{k,i}_m(C^d_n)
\end{align*}

where

\[
\mathit{Bck0}^{k,i}_{m}=\begin{cases}
	r^{k-1,n-1}_m \leftrightarrow r^{k,i}_m & \mbox{if $i=1$} \\
	r^{k,i-1}_m \leftrightarrow r^{k,i}_m & \mbox{otherwise}
\end{cases}
\]

and

\[
\mathit{Bck}^{k,i}_{j,m}=\begin{cases}
r^{k-1,n-1}_m \leftrightarrow r^{k,i}_w \mbox{ with } w = c_{i,j}(m) & \mbox{if $i=1$} \\
r^{k,i-1}_m \leftrightarrow r^{k,i}_w \mbox{ with } w = c_{i,j}(m) & \mbox{otherwise}
\end{cases}
\]

A fixed-depth sorting network for $n$ channels with $d$ layers exists if and only if there is a solution in which all the (unsorted) inputs are sorted, i.e., if $\notsorted(C)=\notsorted^{0,n-1}(C)$ is empty:

\begin{equation}\label{eq:sat_b_depth}
\varphi^b_d(n,d) = \mathit{ValidDepth}(C^d_n) \wedge \mathit{BackwardDepth}(C^d_n) \wedge \mathit{Outputs}^d_n \wedge \mathit{NoUnsortedInputs}_n
\end{equation}

with

\begin{align*}
\mathit{Outputs}^d_n =        & \bigwedge_{m \ne sorted(m)} r^{d,n-1}_m \wedge \bigwedge_{m = sorted(m)} \neg r^{d,n-1}_m \\
\mathit{NoUnsortedInputs}_n = & \bigwedge_{0 \le m < 2^n} \neg r^{0,n-1}_m
\end{align*}

For single-exception fixed-depth sorting networks, we replace $\mathit{NoUnsortedInputs}_n$ with $\mathit{Single}_n$:

\begin{align*}
\mathit{AtMostOneUnsorted}_n = &\bigwedge_{0 \leq m < w < 2^n} (\neg r^{0,n-1}_m \vee \neg r^{0,n-1}_w) \\
\mathit{AtLeastOneUnsorted}_n = &\bigvee_{0 \leq m < 2^n} r^{0,n-1}_m \\
\mathit{Single}_n = &\mathit{AtMostOneUnsorted}_n \wedge \mathit{AtLeastOneUnsorted}_n
\end{align*}

to obtain:

\begin{equation}\label{eq:sat1_depth}
\varphi^1_d(n,d) = \mathit{ValidDepth}(C^d_n) \wedge \mathit{BackwardDepth}(C^d_n) \wedge \mathit{Outputs}^d_n \wedge \mathit{Single}_n
\end{equation}

\section{Results}

In this section, we apply the new family of SAT encodings to three different optimality problems: optimal-size sorting networks, $\epsilon-$halvers and single-exception sorting networks. All the SAT tests are performed with the single-threaded version of the Glucose SAT solver \footnote{http://www.labri.fr/perso/lsimon/glucose}. The software used for these experiments is available at github: \href{https://github.com/jarfo/sort}{https://github.com/jarfo/sort}.

\subsection{Comparison of fixed-size encodings of sorting networks}

For fixed-size formulations, the new forward and backward encodings for the sorting constraint still have the expected exponential size, but they are significantly smaller that the previously proposed encodings \cite{codish2016nine} based on boolean circuit propagation (SCIR). The plain SCIR sorting constraint requires $n s 2^n$ variables, while the proposed SFWD and SBCK sorting constraints need only $s2^n$ variables.

In this first experiment we compare the total number of clauses and variables, including the shared validity constraints, and the solving time of SAT encodings for the optimal-size sorting network problem.

\begin{table}[htb]
\centering\scriptsize\begin{tabular}{|l||r|r|r|r||r|r|r|r||}
\cline{2-9} \multicolumn{1}{c||}{}
&\multicolumn{4}{c||}{optimal-size sorting network (SAT)}      
&\multicolumn{4}{c||}{smaller network (UNSAT)}\\
\hline
encoding & 
$s$ & 
\multicolumn1{c|}{\#clauses} &
\multicolumn1{c|}{\#vars} &
\multicolumn1{c||}{SAT time} &
$s'$&
\multicolumn1{c|}{\#clauses} &
\multicolumn1{c|}{\#vars} &
\multicolumn1{c||}{SAT time} \\
\hline
SCIR & 16  &  108736 &   6510 &    40  & 15  &  101361 &   6048 &   80127\\
SFWD & 16  &   37394 &   2128 &    14  & 15  &   34795 &   1979 &    2591\\
SBCK & 16  &   78763 &   2172 &     6  & 15  &   73496 &   2023 &    4339\\
\hline
\end{tabular}
\caption{SAT-solving size and time for size-$s$ sorting networks on $n=7$ channels. SAT-solving time in seconds (single-threaded glucose solver ). }
\label{fig:sat7}
\end{table}

Table \ref{fig:sat7} clearly shows the important solving-time reduction for optimal-size sorting network problems. The proposed SFWD encoding is 30 times faster that the SCIR encoding proving that there is not sorting network on $7$ channels with $s\le15$ comparators. However, the proposed encodings are still insufficient to give new optimality results with current SAT solvers, and they cannot easily take advantage of fixed network prefixes as the SCIR encoding.

\subsection{Optimal $\epsilon-$halvers}

\begin{figure}[htb]
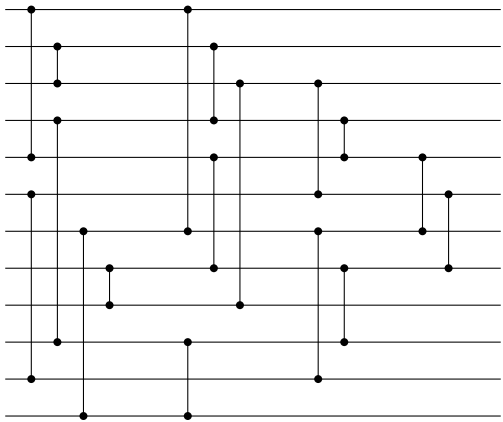

	\centering
    \begin{sortingnetwork}{12}{0.7}
        \nodeconnection{ {0, 4}, {5, 10}}
		\nodeconnection{ {1, 2}, {3, 9}}
		\nodeconnection{ {6, 11}}
		\nodeconnection{ {7, 8}}
		\addtocounter{sncolumncounter}{2}
		\nodeconnection{ {0, 6}, {9, 11}}
		\nodeconnection{ {1, 3}, {4, 7}}
		\nodeconnection{ {2, 8}}
		\addtocounter{sncolumncounter}{2}
		\nodeconnection{ {2, 5}, {6, 10}}
		\nodeconnection{ {3, 4}, {7, 9}}
		\addtocounter{sncolumncounter}{2}
		\nodeconnection{ {4, 6}}
		\nodeconnection{ {5, 7}}
	\end{sortingnetwork}
	\caption{An optimal depth-size $1/4-$halver on $12$ channels with $4$ layers and $17$ comparators.}
	\label{fig:halver12}
\end{figure}

In this experiment we show two examples of the application of the fixed-depth forward encoding (DFWD) to the design of small optimal-depth $\epsilon-$halvers. In the first case we obtain that the optimal depth of a $1/4-$halver for $n=12$ channels is $4$. Including additional size constraints \cite{fonollosa2018joint}, we can also find that the optimal number of comparators for that depth is $17$ (Figure \ref{fig:halver12}).

\begin{figure}[htb]
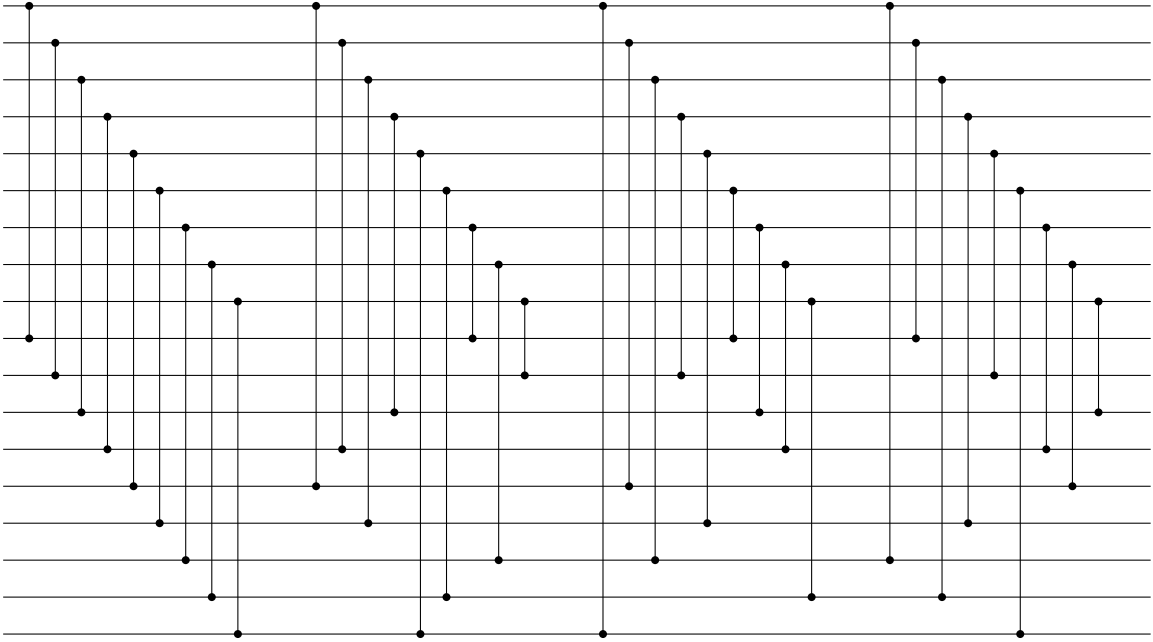

	\centering
	\begin{sortingnetwork}{18}{0.7}
        \nodeconnection{ {0, 9}}
		\nodeconnection{ {1, 10}}
		\nodeconnection{ {2, 11}}
		\nodeconnection{ {3, 12}}
		\nodeconnection{ {4, 13}}
		\nodeconnection{ {5, 14}}
		\nodeconnection{ {6, 15}}
		\nodeconnection{ {7, 16}}
		\nodeconnection{ {8, 17}}
		\addtocounter{sncolumncounter}{2}
		\nodeconnection{ {0, 13}}
		\nodeconnection{ {1, 12}}
		\nodeconnection{ {2, 14}}
		\nodeconnection{ {3, 11}}
		\nodeconnection{ {4, 17}}
		\nodeconnection{ {5, 16}}
		\nodeconnection{ {6, 9}}
		\nodeconnection{ {7, 15}}
		\nodeconnection{ {8, 10}}
		\addtocounter{sncolumncounter}{2}
		\nodeconnection{ {0, 17}}
		\nodeconnection{ {1, 13}}
		\nodeconnection{ {2, 15}}
		\nodeconnection{ {3, 10}}
		\nodeconnection{ {4, 14}}
		\nodeconnection{ {5, 9}}
		\nodeconnection{ {6, 11}}
		\nodeconnection{ {7, 12}}
		\nodeconnection{ {8, 16}}
		\addtocounter{sncolumncounter}{2}
		\nodeconnection{ {0, 15}}
		\nodeconnection{ {1, 9}}
		\nodeconnection{ {2, 16}}
		\nodeconnection{ {3, 14}}
		\nodeconnection{ {4, 10}}
		\nodeconnection{ {5, 17}}
		\nodeconnection{ {6, 12}}
		\nodeconnection{ {7, 13}}
		\nodeconnection{ {8, 11}}
	\end{sortingnetwork}
	\caption{A $1/4-$halver on $18$ channels with $4$ layers and $36$ comparators}
	\label{fig:halver18}
\end{figure}

In the second case we include additional validity constraints to consider only comparators of channels on the upper half with channels on the lower half. Figure \ref{fig:halver18} show the resulting $1/4-$halver on $18$ channels.

\newpage

\subsection{Single-exception sorting networks}

In this experiment we compare single-exception sorting networks with sorting networks in terms of minimum depth and size. Using the fixed-depth backward encoding (DBCK) of single-exception sorting networks $\varphi^1_d(n,d)$ we can obtain optimality results for $n \leq 10$ in a few minutes with current state-of-the-art SAT solvers.

Table \ref{tab:optimald} compares the optimal depth of single-exception sorting networks $D_1(n)$ and sorting networks $D(n)$. Both optimal depths are equal for $n = 4$ and $6 \le n \le 10$, but $D_1(n) = D(n) - 1$ for $n = 5$ and for the trivial cases $n = 2$ and $n = 3$.

\begin{table}[!ht]
	\centering
	\begin{tabular}{|r|c|c|c|c|c|c|c|c|c|}
		\hline 
		$n$      & 2 & 3 & 4 & 5 & 6 & 7 & 8 & 9 & 10 \\
		\hline
		$D_1(n)$ & 0 & 2 & 3 & 4 & 5 & 6 & 6 & 7 &  7 \\
		\hline
		$D(n)$   & 1 & 3 & 3 & 5 & 5 & 6 & 6 & 7 &  7 \\
		\hline
	\end{tabular}
	\caption{Optimal depth of single-exception sorting networks $D_1(n)$ and sorting networks $D(n)$}
	\label{tab:optimald}
\end{table}

We also study the size, and joint size and depth optimization of single-exception sorting networks using the same DBCK encoding with additional size constraints \cite{fonollosa2018joint}. The following tables compare the obtained results for single-exception sorting networks with the previously known results for sorting networks. Note that we can always add a single comparator to a single-exception sorting network to obtain a sorting network. Hence, $D(n) \le D_1(n) + 1$ and $S(n) \le S_1(n) + 1$.

\begin{table}[!ht]
	\centering
	\begin{tabular}{|r|c|c|c|c|c|c|c|c|c|}
		\hline 
		$n$          & 2 & 3 & 4 & 5 &  6 &  7 &  8 &  9 & 10 \\
		\hline
		$S_1(n) \le$ & 0 & 2 & 5 & 8 & 12 & 15 & 20 & 24 & 29 \\
		\hline
		$S_1(n) \ge$ & 0 & 2 & 5 & 8 & 12 & 15 & 18 & 24 & 28 \\
		\hline
		$S(n)$       & 1 & 3 & 5 & 9 & 12 & 16 & 19 & 25 & 29 \\
		\hline
	\end{tabular}
	\caption{Optimal size of single-exception sorting networks $S_1(n)$ and sorting networks $S(n)$}
	\label{tab:optimals}
\end{table}

\begin{table}[!ht]
	\centering
	\begin{tabular}{|r|c|c|c|c|c|c|c|c|c|}
		\hline 
		$n$          &     2 &     3 &     4 &     5 &     6  &      7 &      8 &      9 &            10 \\
		\hline
		$(S,D)_1(n)$ & (0,0) & (2,2) & (5,3) & (8,4) & (12,5) & (15,6) & (20,6) & (24,7) & (29,8),(31,7) \\
		\hline
		$(S,D)(n)$   & (1,1) & (3,3) & (5,3) & (9,5) & (12,5) & (16,6) & (19,6) & (25,7) & (29,8),(31,7) \\
		\hline
	\end{tabular}
	\caption{Optimal (size,depth) combinations of single-exception sorting networks $(S,D)_1(n)$ and sorting networks $(S,D)(n)$ for $n\le10$.}
	\label{tab:optimalsd}
\end{table}

\section{Conclusions}

This paper presents new propositional encodings for the design of optimal comparator networks. In the proposed SAT encodings Boolean variables represent the elements of the set of output vectors after each comparator (or the set of unsorted input vectors), while the clauses encode the effect of each comparator on those sets. The resulting encodings can be easily applied to sorting networks and other comparator networks defined in terms of the set of invalid output vectors such as $\epsilon-$halvers, or the number of unsorted inputs such as single-exception sorting networks.

The experiments show that the proposed encodings can be used to obtain efficient SAT encodings for sorting networks. We also present results of their application to obtain concrete bounds of small $\epsilon-$halvers and single-exception sorting networks.

\section*{References}

\bibliographystyle{plain}
\bibliography{../sorting}

\begin{thebibliography}{10}

\bibitem{abio2013parametric}
Ignasi Ab{\'\i}o, Robert Nieuwenhuis, Albert Oliveras, and Enric
  Rodr{\'\i}guez-Carbonell.
\newblock A parametric approach for smaller and better encodings of cardinality
  constraints.
\newblock In {\em International Conference on Principles and Practice of
  Constraint Programming}, pages 80--96. Springer, 2013.

\bibitem{ajtai1983sorting}
M.~Ajtai, J.~Koml\'{o}s, and E.~Szemer{\'e}di.
\newblock Sorting in c log n parallel steps.
\newblock {\em Combinatorica}, 3(1):1--19, January 1983.

\bibitem{bogdanov2014practical}
Dan Bogdanov, Sven Laur, and Riivo Talviste.
\newblock A practical analysis of oblivious sorting algorithms for secure
  multi-party computation.
\newblock In Karin Bernsmed and Simone Fischer-H{\"u}bner, editors, {\em Secure
  IT Systems}, pages 59--74, Cham, 2014. Springer International Publishing.

\bibitem{bruijn1974}
N.G.~De Bruijn.
\newblock Sorting by means of swappings.
\newblock {\em Discrete Mathematics}, 9(4):333 -- 339, 1974.

\bibitem{bundala2014optimal-depth}
Daniel Bundala, Michael Codish, Lu{\'{\i}}s Cruz{-}Filipe, Peter
  Schneider{-}Kamp, and Jakub Z{\'{a}}vodn{\'{y}}.
\newblock Optimal-depth sorting networks.
\newblock {\em CoRR}, abs/1412.5302, 2014.

\bibitem{chung1987size}
M~Chung and B~Ravikumar.
\newblock On the size of test sets for sorting and related problems.
\newblock In {\em Proc. 1987 International Conference on Parallel Processing},
  1987.

\bibitem{chung1989strong}
Moon~Jung Chung and B.~Ravikumar.
\newblock Strong nondeterministic turing reduction---a technique for proving
  intractability.
\newblock {\em Journal of Computer and System Sciences}, 39(1):2 -- 20, 1989.

\bibitem{codish2016sorting}
Michael Codish, Lu{\'\i}s Cruz-Filipe, Thorsten Ehlers, Mike M{\"u}ller, and
  Peter Schneider-Kamp.
\newblock Sorting networks: To the end and back again.
\newblock {\em Journal of Computer and System Sciences}, 2016.

\bibitem{codish2016nine}
Michael Codish, Lu{\'\i}s Cruz-Filipe, Michael Frank, and Peter Schneider-Kamp.
\newblock Sorting nine inputs requires twenty-five comparisons.
\newblock {\em Journal of Computer and System Sciences}, 82(3):551--563, 2016.

\bibitem{bruijn1983}
N.G. de~Bruijn.
\newblock Sorting arrays by means of swaps.
\newblock {\em Indagationes Mathematicae (Proceedings)}, 86(2):125 -- 132,
  1983.

\bibitem{ehlers2015new}
Thorsten Ehlers and Mike M{\"u}ller.
\newblock New bounds on optimal sorting networks.
\newblock In {\em Conference on Computability in Europe}, pages 167--176.
  Springer, 2015.

\bibitem{fonollosa2018joint}
J.~A.~R. {Fonollosa}.
\newblock {Joint Size and Depth Optimization of Sorting Networks}.
\newblock {\em ArXiv e-prints}, June 2018.

\bibitem{goodrich2014zig}
Michael~T Goodrich.
\newblock Zig-zag sort: A simple deterministic data-oblivious sorting algorithm
  running in o (n log n) time.
\newblock In {\em Proceedings of the forty-sixth annual ACM symposium on Theory
  of computing}, pages 684--693. ACM, 2014.

\bibitem{knuth1998art}
Donald~E. Knuth.
\newblock {\em The Art of Computer Programming, Volume 3: (2Nd Ed.) Sorting and
  Searching}.
\newblock Addison Wesley Longman Publishing Co., Inc., Redwood City, CA, USA,
  1998.

\bibitem{parberry1990single}
Ian Parberry.
\newblock Single-exception sorting networks and the computational complexity of
  optimal sorting network verification.
\newblock {\em Mathematical systems theory}, 23(1):81--93, 1990.

\bibitem{parberry1991computational}
Ian Parberry.
\newblock On the computational complexity of optimal sorting network
  verification.
\newblock In {\em International Conference on Parallel Architectures and
  Languages Europe}, pages 252--269. Springer, 1991.

\end{thebibliography}

\begin{figure}[htb]
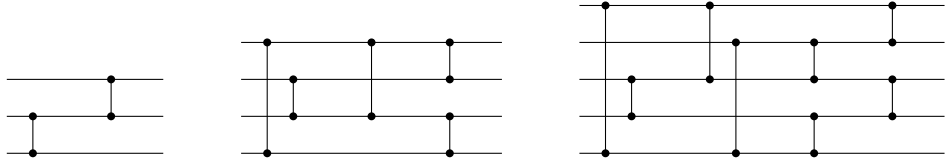

	\centering
	\begin{tabular}{ccc}
		\begin{sortingnetwork}{3}{0.7}
			\nodeconnection{ {1, 2}}
			\addtocounter{sncolumncounter}{2}
			\nodeconnection{ {0, 1}}
		\end{sortingnetwork}
		&
		\begin{sortingnetwork}{4}{0.7}
			\nodeconnection{ {0, 3}}
			\nodeconnection{ {1, 2}}
			\addtocounter{sncolumncounter}{2}
			\nodeconnection{ {0, 2}}
			\addtocounter{sncolumncounter}{2}
			\nodeconnection{ {0, 1}, {2, 3}}
		\end{sortingnetwork}
		&
		\begin{sortingnetwork}{5}{0.7}
			\nodeconnection{ {0, 4}}
			\nodeconnection{ {2, 3}}
			\addtocounter{sncolumncounter}{2}
			\nodeconnection{ {0, 2}}
			\nodeconnection{ {1, 4}}
			\addtocounter{sncolumncounter}{2}
			\nodeconnection{ {1, 2}, {3, 4}}
			\addtocounter{sncolumncounter}{2}
			\nodeconnection{ {0, 1}, {2, 3}}
		\end{sortingnetwork}
	\end{tabular}
	\caption{Optimal single exception sorting networks on $3$, $4$, and $5$ channels}
	\label{fig:optimal5}
\end{figure}

\begin{figure}[htb]
	\centering
	\begin{sortingnetwork}{6}{0.7}
		\nodeconnection{ {1, 4}}
		\nodeconnection{ {3, 5}}
		\addtocounter{sncolumncounter}{2}
		\nodeconnection{ {0, 4}}
		\nodeconnection{ {1, 3}}
		\nodeconnection{ {2, 5}}
		\addtocounter{sncolumncounter}{2}
		\nodeconnection{ {0, 2}, {3, 5}}
		\addtocounter{sncolumncounter}{2}
		\nodeconnection{ {0, 1}, {2, 3}, {4, 5}}
		\addtocounter{sncolumncounter}{2}
		\nodeconnection{ {1, 2}, {3, 4}}
	\end{sortingnetwork}
	\caption{Optimal single exception sorting network on $6$ channels with $5$ layers and $12$ comparators}
	\label{fig:optimal6}
\end{figure}

\begin{figure}[htb]
	\centering
	\begin{sortingnetwork}{7}{0.7}
		\nodeconnection{ {0, 5}}
		\nodeconnection{ {1, 3}, {4, 6}}
		\addtocounter{sncolumncounter}{2}
		\nodeconnection{ {0, 4}}
		\nodeconnection{ {1, 2}, {3, 5}}
		\addtocounter{sncolumncounter}{2}
		\nodeconnection{ {0, 1}, {3, 6}}
		\addtocounter{sncolumncounter}{2}
		\nodeconnection{ {1, 3}, {5, 6}}
		\nodeconnection{ {2, 4}}
		\addtocounter{sncolumncounter}{2}
		\nodeconnection{ {2, 3}, {4, 5}}
		\addtocounter{sncolumncounter}{2}
		\nodeconnection{ {1, 2}, {3, 4}}
	\end{sortingnetwork}
	\caption{Optimal single exception sorting network on $7$ channels with $6$ layers and $15$ comparators}
	\label{fig:optimal7}
\end{figure}

\begin{figure}[htb]
	\centering
	\begin{sortingnetwork}{8}{0.7}
		\nodeconnection{ {0, 1}, {2, 4}, {5, 7}}
		\nodeconnection{ {3, 6}}
		\addtocounter{sncolumncounter}{2}
		\nodeconnection{ {0, 3}, {4, 7}}
		\nodeconnection{ {1, 6}}
		\nodeconnection{ {2, 5}}
		\addtocounter{sncolumncounter}{2}
		\nodeconnection{ {0, 5}}
		\nodeconnection{ {1, 7}}
		\nodeconnection{ {3, 4}}
		\addtocounter{sncolumncounter}{2}
		\nodeconnection{ {0, 2}, {4, 5}, {6, 7}}
		\nodeconnection{ {1, 3}}
		\addtocounter{sncolumncounter}{2}
		\nodeconnection{ {1, 2}, {3, 4}, {5, 6}}
		\addtocounter{sncolumncounter}{2}
		\nodeconnection{ {2, 3}, {4, 5}}
	\end{sortingnetwork}
	\caption{Optimal single exception sorting network on $8$ channels with $6$ layers and $20$ comparators}
	\label{fig:optimal8}
\end{figure}

\begin{figure}[htb]
	\centering
	\begin{sortingnetwork}{9}{0.7}
		\nodeconnection{ {0, 1}, {2, 3}, {4, 5}, {6, 7}}
		\addtocounter{sncolumncounter}{2}
		\nodeconnection{ {0, 2}, {4, 6}}
		\nodeconnection{ {1, 3}, {5, 7}}
		\addtocounter{sncolumncounter}{2}
		\nodeconnection{ {2, 8}}
		\nodeconnection{ {3, 7}}
		\nodeconnection{ {5, 6}}
		\addtocounter{sncolumncounter}{2}
		\nodeconnection{ {0, 4}}
		\nodeconnection{ {1, 6}}
		\nodeconnection{ {2, 5}}
		\nodeconnection{ {3, 8}}
		\addtocounter{sncolumncounter}{2}
		\nodeconnection{ {1, 4}, {6, 8}}
		\nodeconnection{ {3, 5}}
		\addtocounter{sncolumncounter}{2}
		\nodeconnection{ {1, 2}, {3, 4}, {5, 6}, {7, 8}}
		\addtocounter{sncolumncounter}{2}
		\nodeconnection{ {2, 3}, {4, 5}}
	\end{sortingnetwork}
	\caption{Optimal single exception sorting network on $9$ channels with $7$ layers and $24$ comparators}
	\label{fig:optimal9}
\end{figure}

\begin{figure}[htb]
	\centering
	\begin{sortingnetwork}{10}{0.7}
		\nodeconnection{ {0, 4}, {5, 9}}
		\nodeconnection{ {1, 6}}
		\nodeconnection{ {2, 7}}
		\nodeconnection{ {3, 8}}
		\addtocounter{sncolumncounter}{2}
		\nodeconnection{ {0, 1}, {3, 5}, {6, 9}}
		\nodeconnection{ {4, 7}}
		\addtocounter{sncolumncounter}{2}
		\nodeconnection{ {0, 2}, {4, 6}, {8, 9}}
		\nodeconnection{ {1, 5}}
		\addtocounter{sncolumncounter}{2}
		\nodeconnection{ {2, 8}}
		\nodeconnection{ {3, 4}, {7, 9}}
		\addtocounter{sncolumncounter}{2}
		\nodeconnection{ {0, 3}, {5, 7}}
		\nodeconnection{ {1, 2}, {6, 8}}
		\addtocounter{sncolumncounter}{2}
		\nodeconnection{ {1, 3}, {5, 6}, {7, 8}}
		\nodeconnection{ {2, 4}}
		\addtocounter{sncolumncounter}{2}
		\nodeconnection{ {2, 3}, {4, 5}, {6, 7}}
		\addtocounter{sncolumncounter}{2}
		\nodeconnection{ {3, 4}, {5, 6}}
	\end{sortingnetwork}
	\caption{Optimal single exception sorting network on $10$ channels with $8$ layers and $29$ comparators}
	\label{fig:optimal10.8}
\end{figure}

\begin{figure}[htb]
	\centering
	\begin{sortingnetwork}{10}{0.7}
		\nodeconnection{ {0, 6}}
		\nodeconnection{ {1, 7}}
		\nodeconnection{ {2, 5}}
		\nodeconnection{ {3, 8}}
		\nodeconnection{ {4, 9}}
		\addtocounter{sncolumncounter}{2}
		\nodeconnection{ {0, 5}, {7, 8}}
		\nodeconnection{ {1, 3}, {4, 6}}
		\nodeconnection{ {2, 9}}
		\addtocounter{sncolumncounter}{2}
		\nodeconnection{ {0, 4}, {5, 9}}
		\nodeconnection{ {2, 7}}
		\nodeconnection{ {3, 6}}
		\addtocounter{sncolumncounter}{2}
		\nodeconnection{ {0, 3}, {4, 7}}
		\nodeconnection{ {1, 2}, {5, 8}}
		\nodeconnection{ {6, 9}}
		\addtocounter{sncolumncounter}{2}
		\nodeconnection{ {0, 1}, {2, 4}, {6, 7}, {8, 9}}
		\nodeconnection{ {3, 5}}
		\addtocounter{sncolumncounter}{2}
		\nodeconnection{ {1, 2}, {3, 4}, {5, 6}, {7, 8}}
		\addtocounter{sncolumncounter}{2}
		\nodeconnection{ {2, 3}, {4, 5}, {6, 7}}
	\end{sortingnetwork}
	\caption{Optimal single exception sorting network on $10$ channels with $7$ layers and $31$ comparators}
	\label{fig:optimal10.7}
\end{figure}

\end{document}